\documentclass[acmtog]{acmart}

\AtBeginDocument{%
  \providecommand\BibTeX{{%
    \normalfont B\kern-0.5em{\scshape i\kern-0.25em b}\kern-0.8em\TeX}}}

\usepackage{amsmath}
\usepackage[linesnumbered,ruled]{algorithm2e}

\makeatletter
\def\BState{\State\hskip-\ALG@thistlm}
\makeatother

\begin{document}

\title{ML with HE: Privacy Preserving Machine Learning Inferences for Genome Studies}


\author{Şeyma Selcan Mağara}
\affiliation{%
  \institution{Sabanci University}
  \city{Istanbul}
  \country{Turkey}}

\author{Ceren Yıldırım}
\affiliation{%
  \institution{Sabanci University}
  \city{Istanbul}
  \country{Turkey}}
  
\author{Ferhat Yaman}
\affiliation{%
  \institution{North Carolina State University}
  \city{Raleigh}
  \state{North Carolina}
  \country{USA}}
\email{fyaman@ncsu.edu}

\author{Berke Dilekoğlu}
\affiliation{%
  \institution{Sabanci University}
  \city{Istanbul}
  \country{Turkey}}
  
\author{Furkan Reha Tutaş}
\affiliation{%
 \institution{Sabanci University}
  \city{Istanbul}
  \country{Turkey}}

\author{Erdinç Öztürk}
\affiliation{%
  \institution{Sabanci University}
  \city{Istanbul}
  \country{Turkey}}
\email{erdinco@sabanciuniv.edu}

\author{Kamer Kaya}
\affiliation{%
  \institution{Sabanci University}
  \city{Istanbul}
  \country{Turkey}}
\email{kaya@sabanciuniv.edu}

\author{Öznur Taştan}
\affiliation{%
  \institution{Sabanci University}
  \city{Istanbul}
  \country{Turkey}}
\email{otastan@sabanciuniv.edu}

\author{Erkay Savaş}
\affiliation{%
  \institution{Sabanci University}
  \city{Istanbul}
  \country{Turkey}}
\email{erkays@sabanciuniv.edu}

\newcommand\erkays[1]{\textcolor{red}{(\textbf{erkays says :} #1)}}

\newcommand\ceren[1]{\textcolor{orange}{(\textbf{Ceren :} #1)}}

\newcommand\seyma[1]{\textcolor{magenta}{(\textbf{Seyma :} #1)}}

\newcommand\fyaman[1]{\textcolor{blue}{#1}}

\newcommand\furkanreha[1]{\textcolor{brown}{#1}}

\renewcommand{\shortauthors}{Magara, et al.}

\begin{abstract}
Preserving the privacy and security of big data in the context of cloud computing, while maintaining a certain level of efficiency of its processing remains to be a subject, open for improvement. One of the most popular applications epitomizing said concerns is found to be useful in genome analysis. This work proposes a secure multi-label tumor classification method using homomorphic encryption, whereby two different machine learning algorithms, SVM and XGBoost, are used to classify the encrypted genome data of different tumor types.
\end{abstract}



\maketitle

\section{Introduction}
Genome analysis is a popular and rapidly growing area; yet it is crucial to preserve the security and privacy of sensitive genome data while being able to work on it in an efficient manner. Our work is based on iDASH 2020 Competition\footnote{iDASH'20 competition details at \url{http://www.humangenomeprivacy.org/2020/competition-tasks.html}} Track 1, for which the goal was to classify genomes of different tumors with their respective types, such as skin, colon, kidney etc. Due to the high sensitivity of the genomic data, it is crucial to ensure confidentiality. For this purpose we implemented a homomorphic classification model and while doing that, we aimed to achieve a high rate of accuracy in a relatively short amount of time.
 Two different machine learning algorithms were selected to perform this classification task: SVM and XGBoost\footnote{Code is available at \url{https://github.com/SU-CISEC/MLwithHE}}. We initially worked with SVM over the encrypted space since it is useful for classifying models with high dimensional spaces. Furthermore, we implemented an XGBoost algorithm as well since tree based machine learning algorithms have not been widely used in homomorphic applications before. Finally, the accuracies of both algorithms were compared to each other.

\section{Background}
\subsection{Homomorphic Encryption} 

Homomorphic encryption (HE) is an encryption technique that allows computation on encrypted data without decrypting it first. This way the data remains confidential while processing it, hence private data can be outsourced to be processed in untrusted environments \cite{Armknecht2015AGT}. Fully homomorphic encryption (FHE) allows arbitrary computations on encrypted data without the decryption key.
Given encryptions $E({m_1}),\dots, E({m_t})$ of messages ${m_1},\dots, {m_t}$, FHE scheme allows the user to apply the function $f$ on the encrypted messages and obtain a ciphertext which decrypts to $f({m_1}, \dots, {m_t})$ \cite{gentry}. There are also different FHE schemes such as BFV\cite{FV2012}, CKKS\cite{CKKS2017}. BFV scheme utilizes homomorphic operations on integer values while CKKS uses floating point numbers for approximate calculations. 

\subsection{SVM and XGboost}

\subsubsection{SVM}
Support Vector Machine (SVM) is a supervised machine learning algorithm that is used for classification and regression problems. SVM non-linearly maps the input vectors into a high-dimensional feature space and aims to construct a hyperplane that results in the largest margin $w$ between the classes. Given a set of labelled training patterns \(({y_1}, {x_1}), ... , ({y_l}, {x_l}), \; {y_i} \in \{-1, 1\}\), is said to be linearly separable if there exists a vector $w$ and scalar $b$ such that 
\[{y_i} (w \cdot {x_i} + b) \geq 1, \;\; i = 1, ... , l \]
The optimal hyperplane is $w_0 \cdot x + b_0 = 0 $
the unique one which separates the training data with the maximal margin  \cite{cortes_vapnik_1995}.

\subsubsection{XGboost}
XGBoost is a supervised learning algorithm that uses gradient-boosted tree ensembles. Training data is used to construct the model that consists of a set of classification and regression trees (CART). Unlike regular decision trees, CARTs do not contain decision values on the leaves. Instead, each leaf holds a numerical score (i.e. similarity score). Inputs are evaluated on all the trees of the ensemble, which are then classified into one of the leaves. Finally, the numerical scores that we get from each tree are summed up to form a final prediction score. This tree ensemble model that is used to predict the outputs can be shown as
\[ \hat{y_{i}} = \sum_{k=1}^{K} f_{k} (x_{i}), \;\; f_{k} \in F \]
where \(x_{i}\) is the input, \(K\) is the number of the trees in the ensemble, \(F\) is the space of regression trees and each \(f_k\) corresponds to an independent tree structure $q$ and leaf weights $w$  \cite{Chen_2016}.

In the model, each class holds the same number of trees that are used to predict the score of the class for the given input. Once the input is evaluated on the trees of a class, predicted scores from these trees are added up to form a final score for the respective class. The prediction result for the input is the class that holds the highest score value \cite{Chen_2016}.

\subsection{Previous Works} 
Nowadays, machine learning models are used widely in real-life problems. Most of the data used for both training and inference are sensitive information. Thus, preserving privacy and the security of the data has become a priority issue that needs to be solved. Many research conducted on the inference phase such as \cite{6755320} and \cite{8057444}. Recently, \cite{9040596} proposed a secure training method to protect both model information and training data. In this work, it assumed that the model was stored in a secure machine. So we focus on the prediction phase.

Although XGBoost is relatively newer than SVM, it is a commonly used and effective algorithm. Yet, applying HE on XGboost is a challenging task because the prediction phase of the XGBoost algorithm depends on comparison operation. \cite{meng2020privacypreserving} proposed a privacy-preserving XGBoost inference method implemented on Amazon SageMaker. There are also secure and privacy-preserving XGBoost studies that don't use HE. \cite{law2020secure} implemented a solution using hardware enclaves.

\section{Methodology} 
To achieve the most accurate machine learning model for the given genome data, we implemented multiple algorithms with various parameters. Then we evaluated them based on accuracy, speed, and feasibility. Among these, the most proper two models, SVM and XGBoost, are selected to adapt to HE computations. First, we will explain the data set. Then, in the subsections, SVM and XGBoost adaptations are detailed. 

The database of tumor classification task was provided by the IDASH'20 competition. There are total of 2713 unique patient id among 11 different cancer categories in the training data. 

Two types of feature data set were given in the competition. The first one includes the variants of the somatic mutations for all the samples. The second one consists of the copy number states for each tumor type. Copy number states represent deletion and amplification on genes. They are given in 5 different levels where -2 and -1 represent a deletion while 1 and 2 represent an amplification. If none of these are present, it is indicated by 0. We encode our data by using -1 for deletion, 1 for amplification and 0 for neither of these to reduce the complexity.

\subsection{SVM Model}
We implemented one vs. all SVM classification. The algorithm finds optimal hyperplanes that seperate unique class from all other classes in training phase. In the inference phase, confidence values $y_i$ are calculated by dot product of feature vector $X$ and support vector of hyperplane $W_i$ with addition of bias value $b_i$ for every class $i$. 
\[{y_i} = ({W_i} \odot {X} + {b_i}) , \;\; i = 1, ... , s \]
 Feature values $X$ are given encrypted and our model parameters $W$ and $B$ matrices used plaintext in computations. Converting those operations into homomorphic space efficiently requires adjustment of feature size and other homomorphic parameters. We decrease our feature size to minimize inference time while keeping accuracy high. We implemented our homomorphic computations using PALISADE\cite{PALISADE} HE library. Due to floating point numbers in $W$ matrix, we adopted CKKS scheme. Encoding method in CKKS scheme allows to use single instruction multiple data (SIMD) parallelization. Implementation details of homomorphic SVM inference is given in Algorithm~\ref{alg:svm}. Single ciphertext feature vector $X$ and packed weight vector $W$ are multiplied by ciphertext-plaintext homomorphic multiplication using library function. That produces elementwise multiplication results in one packed ciphertext. To calculate dot product, we utilized summation operation in single ciphertext of multiplication results. After that, bias values $B$ added by ciphertext-plaintext homomorphic addition which returns confidence values $Y$ of classification results in encrypted format.
\begin{algorithm}
    \SetKwInOut{Input}{Input}
    \SetKwInOut{Output}{Output}
    \Input{
    $\textbf{X}$: Encrypted feature values \newline
    $W$: Weight Matrix \newline
    $B$: Bias Vector \newline
    $s$: Number of Classes
    }
    \Output{$Y$: Confidence values.}
    \For {$i = 0,1,\ldots,s-1$}
      {
      $W[i] \leftarrow CKKS.PackedPlaintext(W[i])$\;
      $\textbf{t} \leftarrow \textbf{X} * W[i]$ \;
      $\textbf{t} \leftarrow CKKS.Sum(\textbf{t})$\;
      $\textbf{Y[i]} \leftarrow \textbf{t} + B[i]$\;
      }
    \Return $\textbf{Y}$
    \caption{Homomorphic SVM Inference}
    \label{alg:svm}
\end{algorithm}
\subsection{XGBoost Model}
The depth of the tree and the number of trees are two critical parameters that affect the performance. We prefer the depth of the tree to be small since it is directly proportional to the complexity of the model. Also, trees are grown one after another due to the boosting feature of the XGBoost. So, an increase up to a point in the number of trees will enhance accuracy. Considering these facts and evaluating models with various parameters, we selected the depth as two and the number of trees as 128. Hence, the model consists of 128*11 trees of depth 2. 
\begin{figure}
\centering
  \includegraphics[width=0.75\linewidth]{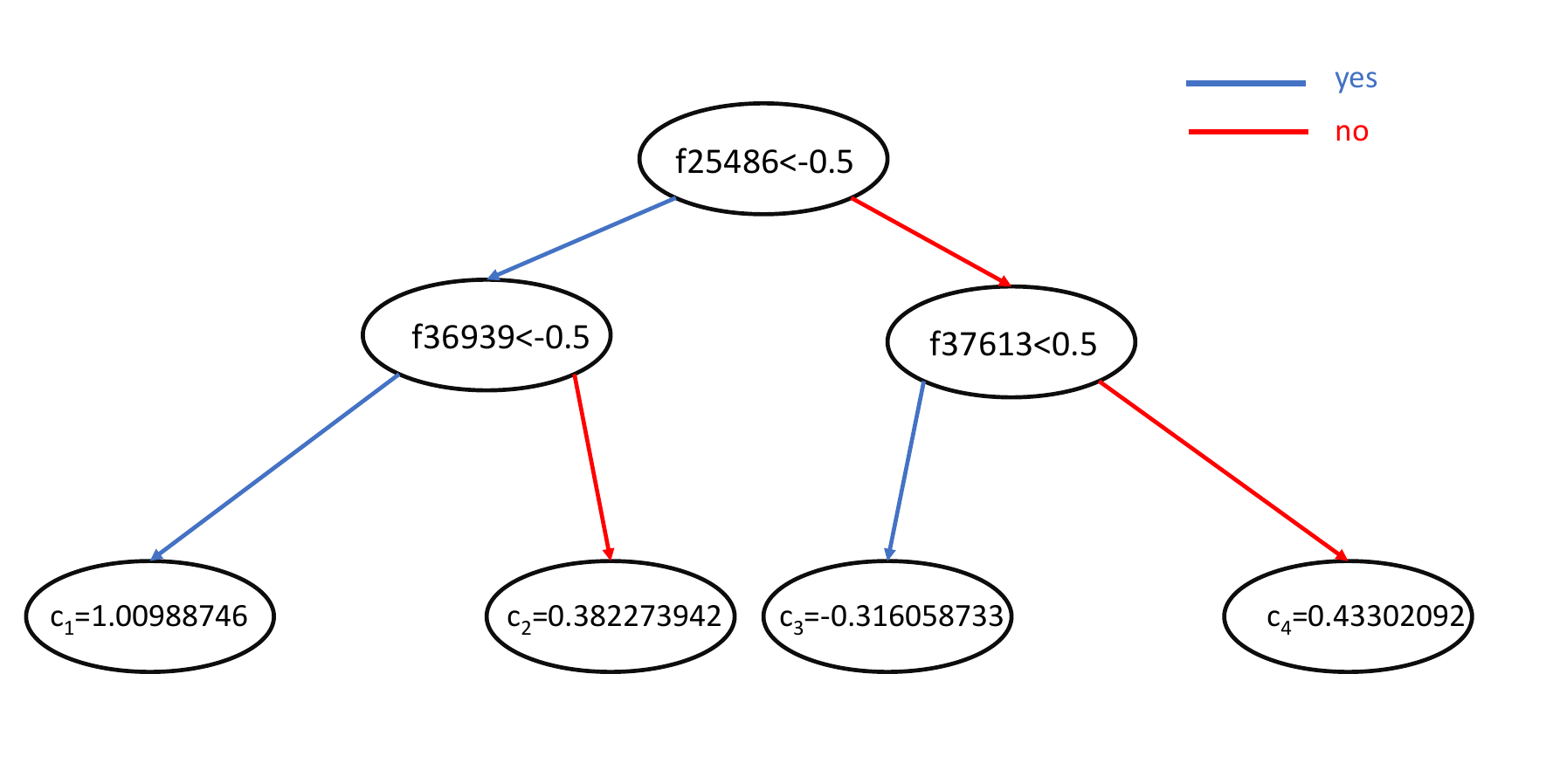}
  \captionof{figure}{A sample tree from our XGBoost model}
  \label{fig:tree}
  \end{figure}
The model searches for optimal split-point for the feature values and leaf scores. For our model, the split points of the nodes are either -0.5 or +0.5. Test data is led into different leaves, and the score of the leaf was added to the corresponding class. A sample tree from the model is given in the \ref{fig:tree}. In this tree, the root node compares the value of the 25486$^{th}$ feature of the test data to $-0.5$. If it is smaller it goes to the left node, otherwise to the right node. The challenging part is comparing the node values with the test data because the data should be encrypted for the sake of privacy. Hence, we decided to convert the comparison operations to logical gate operations and implemented an efficient encoding method similar to one-hot encoding. 
\begin{figure}[h]
  \centering
  \includegraphics[width=.5\linewidth]{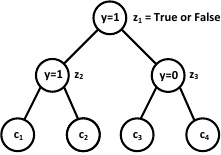}
  \captionof{figure}{Encoded representation of the sample tree in Figure \ref{fig:tree}}
  \label{fig:tree_enc}
\end{figure}  
\begin{table}
\caption{encoding table}
\label{tab:enc_table}
\vspace{-1.5em}
\begin{tabular}{llll}
\toprule
test data value & x2       & x1      & x0      \\
\midrule
-1              & 0        & 0       & 1       \\
0               & 0        & 1       & 0       \\
+1              & 1        & 0       & 0      \\
\bottomrule
\end{tabular}
\end{table}

\begin{table*}[hbt!]
\vspace{-1em}
\centering
\caption{Performance Table: Execution times (Key generation, encryption, computation, decryption and total timing) and MicroAUCs}
\label{tab:results}
\scalebox{1}{
\begin{tabular}{ccccccccl}
         &          &       & KeyGen & Enc             & Comp        & Dec         & EndtoEnd    & microAUC\\\hline
SVM      & Palisade &       & 0.06s   & 0.078s           & 2.748s       & 1.029s       & 3.915s     & 0.96\\\hline
XGBoost  & SEAL     & BFV   & 0.31s   & 1.31s            & 5.106s       & 1.671s           & 8.397s  &   0.981 \\\hline
XGBoost  & Palisade & HEAAN & 0.257s  & 4.984s           & 10.501s      & 10.218s          & 25.96s   &  0.981 \\\hline
XGBoost(Encrypted Model) & Palisade & HEAAN & 0.07s   & 5.799s           & 16.664s      & 1.399s           & 23.932s   &   0.981   
\end{tabular}}
\end{table*}
Since the split points of the nodes can take only two values, 1 bit is sufficient to encode the node values. Therefore, split values equal to 0.5 are replaced with 0 and -0.5 is replaced with 1. We show the encoded split-points with letter $y$ in this paper.
The test data is represented by only 3 possible values: -1, 0, 1 Therefore, 3 bits are needed to represent test data with one-hot encoding, namely $x_0$, $x_1$ and $x_2$. If, value of a feature is -1 then $x_2$, $x_1$ and $x_0$ values are mapped to 0, 0, 1. Full encoding is given in Table~\ref{tab:enc_table}.

We used Karnaugh maps (kmap) in order to achieve an optimized boolean expression that has the same behaviour as the comparison operation. As a result, we expressed a single comparison as \[ \neg x_2 \land (\neg y \lor x_0)\] Notice that we were able to formulate it without the $x_1$ term. This significantly reduces the complexity.
\begin{figure}[]
  \centering
  \includegraphics[width=0.45\linewidth]{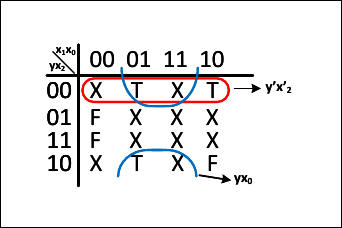}
  \caption{kmap is used to obtain optimized boolean expression. X's show don't care conditions}
  \label{fig:kmap}
\end{figure}
Since we are working with BFV and CKKS HE schemes, we are limited to arithmetical operations. However all the terms are in the binary field. Thus we can convert this boolean expression into an arithmetical expression effortlessly: 
\[z = (1-x_0) *(x_2*(y-1)-y)+1 \]
Using this formula we can compare the value of a feature with the value of a node. If it is smaller result will be 1, otherwise 0. Remember that, prediction data is encrypted. Thus, the result of the comparison is also encrypted. We need to reach the correct leaf node without decrypting it. To solve this issue, we took advantage of node values being boolean.

In complete binary trees of depth 2, there are four possible paths. As shown in the Fig.~\ref{fig:tree_enc}, truth value of a node i is shown with $z_i$ and the score of the leaves with $c_i$. Then, the values of paths can be written as:
\begin{align*}
p_1 &= z_1*z_2*c_1     
p_2 = z_1*(1-z_2)*c_2 \\
p_3 &= (1-z_1)*z_3*c_3      
p_4 = (1-z_1)*(1-z_3)*c_4 
\end{align*}

Note that, only one of these, the correct path, can be nonzero at a time. Others will be zero. So, summation of the all $p_i$'s gives the score of a single tree.  Simplified score expression will be :\[score = ({z_1} - 1) * {l_3} * {z_3} + ({z_2} * {l_1} + {l_2}) * {z_1} + {l_4}\]

\section{Implementation}
\begin{algorithm}
    \SetKwInOut{Input}{Input}
    \SetKwInOut{Output}{Output}
    \Input{$X_i = [x_{i2}, x_{i1}, x_{i0}]$: Encoded feature values of all trees appended in order. \newline
    $y_i$: Encoded node values of all trees appended in order. \newline
    for $i = 0,1,\ldots,m-1 $ \newline
    $m$: number of genetic information inputs.}
    \Output{$z$: Comparison result of a single node value with the feature value it holds.}
    \For {$i = 0,1,\ldots,m-1$}
      {
        $z[i] \leftarrow (1 - {x_{i0}}) * ({x_{i2}} * ({y_i} - 1) - {y_i}) + 1$
      }
      
    \Return z

    \caption{Comparing Feature Values With Leaf Values}
    \label{algo:comp}
\end{algorithm}
\begin{algorithm}
    \SetKwInOut{Input}{Input}
    \SetKwInOut{Output}{Output}
    \Input{$(z_1,z_2,z_3)$: Comparison results for each inner node.\newline
    $L = \{l_1, l_2, l_3, l_4\}$: List of leaf values. \newline
    $m$ : Number of genetic information inputs.
    }
    \Output{$R$: List of final scores of each genetic information input.}
    $R\leftarrow{(0)^{1xm}}$
    
    \For {$i = 0,1,\ldots,m-1$}
        {
        $R[i]\leftarrow{({z_1} - 1) * {l_3} * {z_3} + ({z_2} * {l_1} + {l_2}) * {z_1} + {l_4}}$
        }
    \Return $R$
    \caption{Calculating Tree Score}
    \label{algo:calc_tree}
\end{algorithm}
The basic outline of how this detection system works is given as below:
\begin{enumerate}
    \item The model is not encrypted. On the other hand, the genomic input data is given in encrypted form. Hence, the client sends their data to the server and computation is done on the cloud side. 
    \item Once the input data is received, the cloud server computes the decision values for each node of every tree in the model. This is done by comparing the numeric values given in the inner node with the specified feature value. By using the formula given in Algorithm \ref{algo:comp}, the result is either 0 or 1, depending on whether the condition given for the node is satisfied or not. This operation is done homomorphically since the feature values are encrypted.
    \item For each tree, the server stores the output of its inner nodes. Consequently the information for which leaf node the input is classified into is stored in encrypted format. Using the formula given in Algorithm \ref{algo:calc_tree}, the server creates an array which stores the predicted leaf node's value for all the trees in the model. This array of size $l x k$, where $l$ stands for the number of classes and $k$ stands for the number of trees for each class.
    \item Finally for each class, the tree scores are summed up in logarithmic time. After the summation, the final score for each class $s = 0,1,\ldots,l-1$ is stored in the $(s * k)$ th cell of the array in encrypted form. Once the client gets this array and decrypts it, the class with the highest score will be the model's prediction for the client's input.
\end{enumerate}
There are three inner nodes in each tree. Hence, compareNode function is called three times for each inner node and the comparison values of the root node, left child and right child are stored in ${z_1}, {z_2}$ and ${z_3}$ respectively.
Once we have the comparison results for every node, calculateTree function is called. This function returns an array that holds the leaf score of the predicted output for each tree. 

\section{Results} 
\begin{figure}[h]
  \centering
  \includegraphics[width=.9\linewidth]{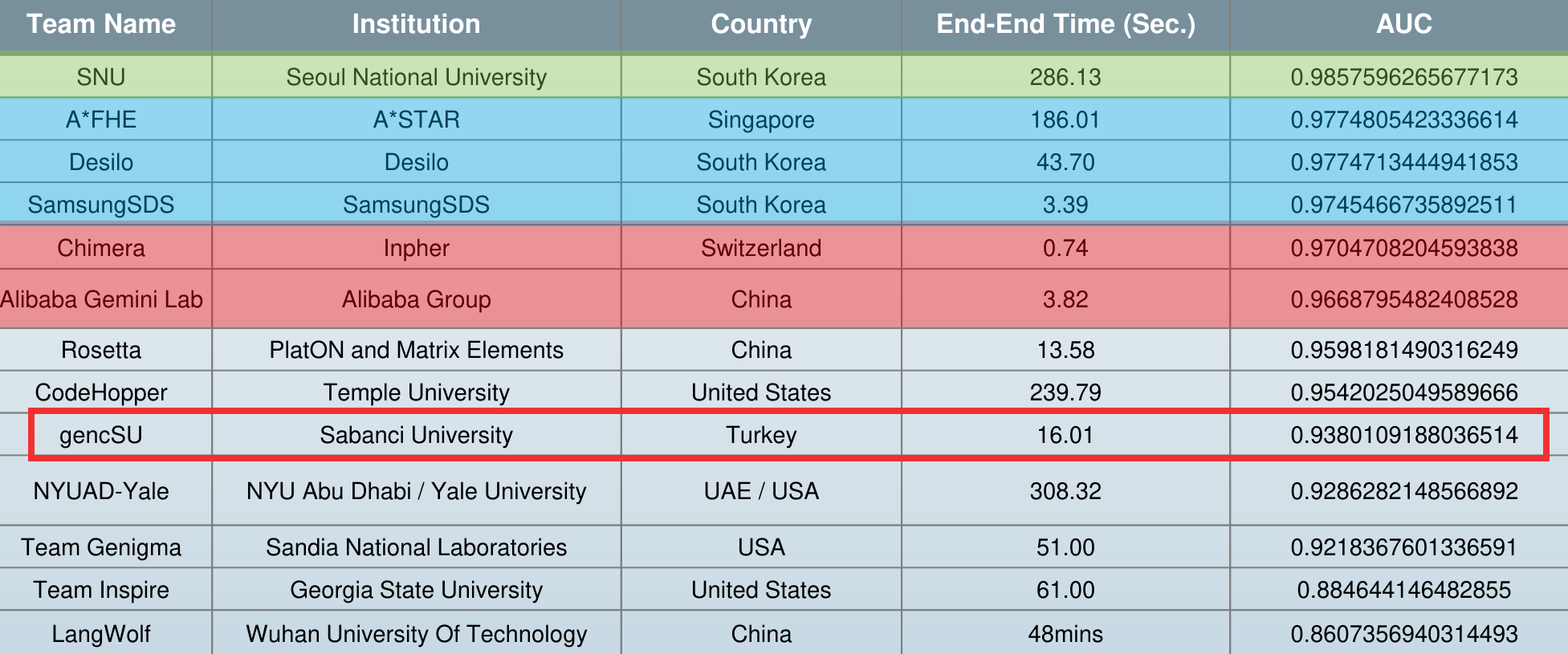}
  \caption{IDASH'20 End-End time sorted rankings Accessed 2 August 2021, <http://www.humangenomeprivacy.org/2020/agenda.html>}
  \label{fig:comp_result}
\end{figure}

We participated in the IDASH’20 competition with the SVM model, and achieved remarkable results as seen in the Figure~\ref{fig:comp_result}. In addition, an XGBoost model is developed using different HE libraries(SEAL\cite{sealcrypto} and Palisade) and schemes. The comparison between XGBoost and SVM models is shown in Table~\ref{tab:results}. Based on the measurements, the XGBoost model is superior to SVM in terms of accuracy while being slower. This is expected because XGBoost has a higher multiplication depth, and it uses larger data for the prediction phase. 
Since the competition ranked the participants using a separate data set, a direct comparison is not possible. However, it is possible to make an inference about the performance of the XGBoost model by comparing it to other participants' results. We have achieved a significant increase in the accuracy with the XGBoost model. While the speed of the XGBoost model is lower than that of the SVM model's, it still compares well to other participants' timing results. We believe this improvement over the accuracy makes up for the increase in end-to-end timing. 
\vspace{-2em}
\section{Conclusion}
In this paper, we proposed privacy-preserving SVM and XGBoost classification models using HE. XGBoost inference phase is adapted for HE with an efficient encoding method that speeds up the costly secure comparison operation. Furthermore, we implemented an efficient and accurate homomorphic SVM inference model. Timing and accuracy measurements showed that the XGBoost algorithm with HE is highly applicable and efficient. In addition, these models and the experiments confirmed the effectiveness of HE on machine learning models. 

\section{Acknowledgement}
C. Yıldırım, Dr. Öztürk and Dr. Savaş are supported by TUBITAK under grant number 118E725.

\bibliographystyle{ACM-Reference-Format}
\bibliography{sample-base}


\begin{thebibliography}{13}


\ifx \showCODEN    \undefined \def \showCODEN     #1{\unskip}     \fi
\ifx \showDOI      \undefined \def \showDOI       #1{#1}\fi
\ifx \showISBNx    \undefined \def \showISBNx     #1{\unskip}     \fi
\ifx \showISBNxiii \undefined \def \showISBNxiii  #1{\unskip}     \fi
\ifx \showISSN     \undefined \def \showISSN      #1{\unskip}     \fi
\ifx \showLCCN     \undefined \def \showLCCN      #1{\unskip}     \fi
\ifx \shownote     \undefined \def \shownote      #1{#1}          \fi
\ifx \showarticletitle \undefined \def \showarticletitle #1{#1}   \fi
\ifx \showURL      \undefined \def \showURL       {\relax}        \fi
\providecommand\bibfield[2]{#2}
\providecommand\bibinfo[2]{#2}
\providecommand\natexlab[1]{#1}
\providecommand\showeprint[2][]{arXiv:#2}

\bibitem[\protect\citeauthoryear{??}{PAL}{2021}]%
        {PALISADE}
 \bibinfo{year}{2021}\natexlab{}.
\newblock \bibinfo{title}{{PALISADE} {L}attice {C}ryptography {L}ibrary
  (release 1.11.2)}.
\newblock \bibinfo{howpublished}{\url{https://palisade-crypto.org/}}.
\newblock


\bibitem[\protect\citeauthoryear{Armknecht, Boyd, Carr, Gj{\o}steen,
  J{\"a}schke, Reuter, and Strand}{Armknecht et~al\mbox{.}}{2015}]%
        {Armknecht2015AGT}
\bibfield{author}{\bibinfo{person}{Frederik Armknecht}, \bibinfo{person}{C.
  Boyd}, \bibinfo{person}{Christopher Carr}, \bibinfo{person}{K. Gj{\o}steen},
  \bibinfo{person}{Angela J{\"a}schke}, \bibinfo{person}{Christian~A. Reuter},
  {and} \bibinfo{person}{Martin Strand}.} \bibinfo{year}{2015}\natexlab{}.
\newblock \showarticletitle{A Guide to Fully Homomorphic Encryption}.
\newblock \bibinfo{journal}{\emph{IACR Cryptol. ePrint Arch.}}
  \bibinfo{volume}{2015} (\bibinfo{year}{2015}), \bibinfo{pages}{1192}.
\newblock


\bibitem[\protect\citeauthoryear{Chen and Guestrin}{Chen and Guestrin}{2016}]%
        {Chen_2016}
\bibfield{author}{\bibinfo{person}{Tianqi Chen} {and} \bibinfo{person}{Carlos
  Guestrin}.} \bibinfo{year}{2016}\natexlab{}.
\newblock \showarticletitle{XGBoost}.
\newblock \bibinfo{journal}{\emph{Proceedings of the 22nd ACM SIGKDD
  International Conference on Knowledge Discovery and Data Mining}}
  (\bibinfo{date}{Aug} \bibinfo{year}{2016}).
\newblock
\showISBNx{9781450342322}
\urldef\tempurl%
\url{https://doi.org/10.1145/2939672.2939785}
\showDOI{\tempurl}


\bibitem[\protect\citeauthoryear{Cheon, Kim, Kim, and Song}{Cheon
  et~al\mbox{.}}{2017}]%
        {CKKS2017}
\bibfield{author}{\bibinfo{person}{Jung~Hee Cheon}, \bibinfo{person}{Andrey
  Kim}, \bibinfo{person}{Miran Kim}, {and} \bibinfo{person}{Yongsoo Song}.}
  \bibinfo{year}{2017}\natexlab{}.
\newblock \showarticletitle{Homomorphic Encryption for Arithmetic of
  Approximate Numbers}. In \bibinfo{booktitle}{\emph{Advances in Cryptology --
  ASIACRYPT 2017}}, \bibfield{editor}{\bibinfo{person}{Tsuyoshi Takagi} {and}
  \bibinfo{person}{Thomas Peyrin}} (Eds.). \bibinfo{publisher}{Springer
  International Publishing}, \bibinfo{address}{Cham},
  \bibinfo{pages}{409--437}.
\newblock
\showISBNx{978-3-319-70694-8}


\bibitem[\protect\citeauthoryear{Cortes and Vapnik}{Cortes and Vapnik}{1995}]%
        {cortes_vapnik_1995}
\bibfield{author}{\bibinfo{person}{Corinna Cortes} {and}
  \bibinfo{person}{Vladimir Vapnik}.} \bibinfo{year}{1995}\natexlab{}.
\newblock \showarticletitle{Support-vector networks}.
\newblock \bibinfo{journal}{\emph{Machine Learning}} \bibinfo{volume}{20},
  \bibinfo{number}{3} (\bibinfo{year}{1995}), \bibinfo{pages}{273–297}.
\newblock
\urldef\tempurl%
\url{https://doi.org/10.1007/bf00994018}
\showDOI{\tempurl}


\bibitem[\protect\citeauthoryear{Fan and Vercauteren}{Fan and
  Vercauteren}{2012}]%
        {FV2012}
\bibfield{author}{\bibinfo{person}{Junfeng Fan} {and} \bibinfo{person}{F.
  Vercauteren}.} \bibinfo{year}{2012}\natexlab{}.
\newblock \showarticletitle{Somewhat Practical Fully Homomorphic Encryption}.
\newblock \bibinfo{journal}{\emph{IACR Cryptol. ePrint Arch.}}
  \bibinfo{volume}{2012} (\bibinfo{year}{2012}), \bibinfo{pages}{144}.
\newblock


\bibitem[\protect\citeauthoryear{Gentry}{Gentry}{2009}]%
        {gentry}
\bibfield{author}{\bibinfo{person}{Craig Gentry}.}
  \bibinfo{year}{2009}\natexlab{}.
\newblock \emph{\bibinfo{title}{A fully homomorphic encryption scheme}}.
\newblock \bibinfo{thesistype}{Ph.D. Dissertation}.
\newblock


\bibitem[\protect\citeauthoryear{Law, Leung, Poddar, Popa, Shi, Sima, Yu,
  Zhang, and Zheng}{Law et~al\mbox{.}}{2020}]%
        {law2020secure}
\bibfield{author}{\bibinfo{person}{Andrew Law}, \bibinfo{person}{Chester
  Leung}, \bibinfo{person}{Rishabh Poddar}, \bibinfo{person}{Raluca~Ada Popa},
  \bibinfo{person}{Chenyu Shi}, \bibinfo{person}{Octavian Sima},
  \bibinfo{person}{Chaofan Yu}, \bibinfo{person}{Xingmeng Zhang}, {and}
  \bibinfo{person}{Wenting Zheng}.} \bibinfo{year}{2020}\natexlab{}.
\newblock \bibinfo{title}{Secure Collaborative Training and Inference for
  XGBoost}.
\newblock
\newblock
\showeprint[arxiv]{2010.02524}~[cs.CR]


\bibitem[\protect\citeauthoryear{Meng and Feigenbaum}{Meng and
  Feigenbaum}{2020}]%
        {meng2020privacypreserving}
\bibfield{author}{\bibinfo{person}{Xianrui Meng} {and} \bibinfo{person}{Joan
  Feigenbaum}.} \bibinfo{year}{2020}\natexlab{}.
\newblock \bibinfo{title}{Privacy-Preserving XGBoost Inference}.
\newblock
\newblock
\showeprint[arxiv]{2011.04789}~[cs.CR]


\bibitem[\protect\citeauthoryear{Omer, Gao, and Sayed}{Omer
  et~al\mbox{.}}{2016}]%
        {8057444}
\bibfield{author}{\bibinfo{person}{Mohammed~Z. Omer}, \bibinfo{person}{Hui
  Gao}, {and} \bibinfo{person}{Faisal Sayed}.} \bibinfo{year}{2016}\natexlab{}.
\newblock \showarticletitle{Privacy Preserving in Distributed SVM Data Mining
  on Vertical Partitioned Data}. In \bibinfo{booktitle}{\emph{2016 3rd
  International Conference on Soft Computing Machine Intelligence (ISCMI)}}.
  \bibinfo{pages}{84--89}.
\newblock
\urldef\tempurl%
\url{https://doi.org/10.1109/ISCMI.2016.40}
\showDOI{\tempurl}


\bibitem[\protect\citeauthoryear{Park, Byun, Lee, Cheon, and Lee}{Park
  et~al\mbox{.}}{2020}]%
        {9040596}
\bibfield{author}{\bibinfo{person}{Saerom Park}, \bibinfo{person}{Junyoung
  Byun}, \bibinfo{person}{Joohee Lee}, \bibinfo{person}{Jung~Hee Cheon}, {and}
  \bibinfo{person}{Jaewook Lee}.} \bibinfo{year}{2020}\natexlab{}.
\newblock \showarticletitle{HE-Friendly Algorithm for Privacy-Preserving SVM
  Training}.
\newblock \bibinfo{journal}{\emph{IEEE Access}}  \bibinfo{volume}{8}
  (\bibinfo{year}{2020}), \bibinfo{pages}{57414--57425}.
\newblock
\urldef\tempurl%
\url{https://doi.org/10.1109/ACCESS.2020.2981818}
\showDOI{\tempurl}


\bibitem[\protect\citeauthoryear{SEAL}{SEAL}{2020}]%
        {sealcrypto}
SEAL \bibinfo{year}{2020}\natexlab{}.
\newblock \bibinfo{title}{{M}icrosoft {SEAL} (release 3.6)}.
\newblock \bibinfo{howpublished}{\url{https://github.com/Microsoft/SEAL}}.
\newblock
\newblock
\shownote{Microsoft Research, Redmond, WA.}


\bibitem[\protect\citeauthoryear{Teo, Han, and Lee}{Teo et~al\mbox{.}}{2013}]%
        {6755320}
\bibfield{author}{\bibinfo{person}{Sin~G. Teo}, \bibinfo{person}{Shuguo Han},
  {and} \bibinfo{person}{Vincent~C.S. Lee}.} \bibinfo{year}{2013}\natexlab{}.
\newblock \showarticletitle{Privacy Preserving Support Vector Machine Using
  Non-linear Kernels on Hadoop Mahout}. In \bibinfo{booktitle}{\emph{2013 IEEE
  16th International Conference on Computational Science and Engineering}}.
  \bibinfo{pages}{941--948}.
\newblock
\urldef\tempurl%
\url{https://doi.org/10.1109/CSE.2013.200}
\showDOI{\tempurl}


\end{thebibliography}

\appendix

\end{document}